\documentclass[preprint,11pt,showkeys,showpacs,nofootinbib,floatfix]{revtex4}
\usepackage{amssymb}
\usepackage{amsfonts}
\usepackage{mathrsfs}
\usepackage{amsmath}
\usepackage{graphicx}
\usepackage{subfigure}
\usepackage[
            pdfstartview=FitH,
            bookmarksnumbered=true,
            bookmarksopen=true,
            colorlinks,
            pdfborder=001,
            linkcolor=blue,
            anchorcolor=green,
            citecolor=blue
            ]{hyperref}
\usepackage[toc,page,title,titletoc,header]{appendix}
\usepackage{pdfpages}
\usepackage{booktabs}
\usepackage{multirow}
\usepackage{cases}
\makeatletter

\newcommand{\Rmnum}[1]{\expandafter\@slowromancap\romannumeral #1@}
\makeatother

\begin{document}
\title{Holographic paramagnetism-ferromagnetism phase transition with the nonlinear electrodynamics}
\author{Cheng-Yuan Zhang$^1$}
\author{Ya-Bo Wu$^1$}
\thanks{Corresponding author\\
E-mail address:ybwu61@163.com}
\author{Ya-Nan Zhang$^1$}
\author{Huan-Yu Wang$^1$}
\author{Meng-Meng Wu$^1$}
\affiliation{$^1$ Department of physics, Liaoning Normal University, Dalian, 116029, China}
%
\begin{abstract}
In the probe limit, we investigate the nonlinear electrodynamical effects of the both exponential form and the logarithmic form on the holographic paramagnetism-ferromagnetism phase transition in the background of a Schwarzschild-AdS black hole spacetime. Moreover, by comparing the exponential form of nonlinear electrodynamics with the logarithmic form of nonlinear electrodynamics and the Born-Infeld nonlinear electrodynamics which has been presented in Ref.~\cite{Wu:2016uyj}, we find that the higher nonlinear electrodynamics correction makes the critical temperature smaller and the magnetic moment harder form in the case without external field. Furthermore, the increase of nonlinear parameter $b$ will result in extending the period of the external magnetic field. Especially, the effect of the exponential form of nonlinear electrodynamics on the periodicity of hysteresis loop is more noticeable.
\end{abstract}

\pacs{11.25.Tq, 04.70.Bw, 11.27.+d, 75.10.-b}

\keywords{AdS/CFT correspondence, Holographic ferromagnetism, Nonlinear electrodynamics}
\maketitle

\section{Introduction}
In the early part of the last century, the phenomenon of superconductivity that the electrical resistivity of a material suddenly drops to zero below a critical temperature $T_{c}$ was discovered and then the most successful microscopic theory of superconductivity was proposed by Bardeen, Cooper and Schrieffer(BCS) to describe various properties of usual (low
temperature) superconducting materials with great accuracy. However, in the modern condensed matter physics, some materials of significant theoretical and practical interest, such as the high-temperature cuprates and heavy fermion compounds, are beyond BCS theory. Fortunately, the anti-de Sitter(AdS)/conformal field theory(CFT) correspondence~\cite{Maldacena:1997re,Gubser:1998bc,Witten:1998qj,Witten:1998zw} provides a window into the strongly coupled condensed matter system, especially the construction of the holographic superconductor, by relating a weak coupling gravity theory in a d-dimensional AdS spacetime to a strong CFT on the (d-1)-dimensional boundary. It was suggested that the instability of the bulk black hole corresponds to a second order phase transition from normal state to superconducting state which brings the spontaneous U(1) symmetry breaking~\cite{Gubser:2008px}. The authors of Ref.~\cite{Hartnoll:2008vx} introduced the first superconductor can indeed be reproduced in this simple model.
And following this, a variety of the holographic dual models have been shown in Refs.~\cite{Hartnoll:2008kx,Herzog:2009xv,Horowitz:2010gk,Nie:2013sda,Cai:2015cya,Ling:2014laa,Mansoori:2016zbp,Kuang:2016edj,Lu:2013tza,Lai:2016yma} .

Recently, some efforts have been made to generalize the AdS/CFT correspondence to magnetism. The authors of Ref.~\cite{Cai:2014oca} realized the holographic description of the paramagnetism-ferromagnetism phase transition in a dyonic Reissner-Nordstr\"{o}m-AdS black brane. In that model, the magnetic moment is realized by the condensation of a real antisymmetric tensor field which couples to the background gauge field strength in the bulk. In the case without an external magnetic field, the time reversal symmetry is spontaneously broken, and the spontaneous magnetization happens in low temperatures. the critical exponents are in agreement with the ones from mean field theory. In the case of a nonzero magnetic field, the model realizes the hysteresis loop of the single magnetic domain, and the magnetic susceptibility satisfies the Curie-Weiss law. Obviously, this model in Ref.~\cite{Cai:2014oca} gives a good starting point to explore more complicated magnetic phenomena and quantum phase transitions. Since then, a large number of the holographic dual models have been constructed and some interesting behaviors have been found, for reviews, see Refs.~\cite{Cai:2014jta,Cai:2014dza,Yokoi:2015qba,Cai:2015bsa,Cai:2015jta,Cai:2015wfa,Cai:2016iim,Zhang:2016gwv} and references therein.

All of the above mentioned models are carried out in the framework of usual Maxwell electrodynamics. However, besides the conventional framework of Maxwell electrodynamics, there is always a provision for non-linear electrodynamics, which correspond to the high derivative corrections to the gauge fields in various aspects of gravity theories. Moreover, it turned out that the higher derivative corrections of the gauge field carries more plentiful information than the usual Maxwell electrodynamics~\cite{Anninos:2008sj,Hendi:2010zza,Miskovic:2010ui,Miskovic:2010ey,Gurtug:2010dr,Shabad:2011hf}, and has been a focus for these years since most physical systems are inherently nonlinear to some extent. Among the various theories with non-linear electrodynamics, the Born-Infeld(BI) theory~\cite{Born:1934gh} has attained renewed attentions due to its several remarkable features. One of the interesting properties of the BI theory is that the electric field is regular for a point-like particle. The regular BI theory with finite energy gives the non-singular solutions of the field equations. In fact the BI electrodynamics is the only non-linear electrodynamic theory with a sensible weak field limit~\cite{Boillat:1970gw}. Another fascinating feature of the BI theory is that it remains invariant under electromagnetic duality. Therefore, considering Born-Infeld electrodynamics, Jing and Chen firstly introduced holographic dual model and observed that the nonlinear Born-Infeld corrections make it harder for the scalar condensation to form~\cite{Jing:2010zp}. Subsequently, some rich physics in the phase transition of the holographic superconductor with Born-Infeld electrodynamics in Gauss-Bonnet gravity has been observed~\cite{Jing:2010cx}. Along this direction, there have been accumulated interest to study various holographic dual models with the nonlinear electrodynamics~\cite{Jing:2011vz,Pan:2011vi,Jing:2015nqv,Sheykhi:2016kqh,Gangopadhyay:2012np,Lee:2012qn,Roychowdhury:2012hp,Gangopadhyay:2012am,
Yao:2013sha,Yao:2014fwa,Lai:2015rva,Ghorai:2015wft,Sheykhi:2015mxb}. At the same time, similar to the case of Born~-~Infeld nonlinear electrodynamics, other types of nonlinear electrodynamics in the context of gravitational field have been introduced, which can also remove the divergence arising in Maxwell theory at the origin. Two well known nonlinear Lagrangian for electrodynamics are logarithmic(LEN)~\cite{Soleng:1995kn,Jing:2012dj} and exponential(ENE)~\cite{Hendi:2012zz,Hendi:2013mka,Yao:2016ils} Lagrangian. The authors of Ref.~\cite{Zhao:2012cn} observed that the exponential form of nonlinear electrodynamics has stronger effect on the condensation formation and conductivity for the holographic conductors in the backgrounds of AdS black hole by considering three types of typical nonlinear electrodynamics. However, in the AdS soliton background the critical chemical potentials are independent of the explicit form of the nonlinear electrodynamics, i.e., the ENE, BINE and LNE correction do not effect on the critical potentials. So based on the research about holographic superconductor with the nonlinear electrodynamics, we have studied the effect of BI coupling parameter on the paramagnetism-ferromagnetism phase transition~\cite{Wu:2016uyj}. And now it is interesting to investigate how the other two types of nonlinear electrodynamics influence the paramagnetism-ferromagnetism phase transition.

The structure of this work is as follows. In section \Rmnum{2}, we introduce the basic field equations of holographic ferromagnetism model with ENE and LEN in the Schwarzschild-AdS black hole which have not been studied as far as we know, and compare them with the BINE holographic paramagnetism model. In section \Rmnum{3} by the numerical method we obtain the critical temperature and study the magnetic moment in the presence of the three kinds of typical nonlinear electrodynamics. Magnetic susceptibility density and hysteresis loop will be shown in section \Rmnum{4}. Finally in the last section we will include our summary and discussion.

\section{Holographic model}
In this paper, the model we are considering is just general relativity with a negative cosmological constant $\Lambda=-3/L^2$, a U(1) field $A_{\mu}$ and a massive 2-form field $M_{\mu\nu}$ in 4-dimension space-time. The ghost free action
\begin{align}\label{action1}
&S=\frac1{2\kappa^2}\int d^4 x\sqrt{-g}(L_1+\lambda^2 L_{2})£¬\\
&L_{1}=R+6/L^2+L(F),\\
&L_{2}=-\frac{1}{12}(d M)^2-\frac{m^2}{4}M_{\mu\nu}M^{\mu\nu}-\frac{1}{2}M^{\mu\nu}F_{\mu\nu}-\frac{J}{8}V(M)
\end{align}
where $d M$ is the exterior differential of 2-form field $M_{\mu\nu}$, $m^2$ is the squared mass of 2-form field $M_{\mu\nu}$ being greater than zero (see Ref.~\cite{Cai:2015bsa} for detail), $\lambda$ and $J$ are two real model parameters with $J<0$ for producing the spontaneous magnetization, $\lambda^2$ characterizes the back reaction of the 2-form field $M_{\mu\nu}$ to the background geometry and to the Maxwell field strength, and $V(M)$ is a nonlinear potential of the 2-form field describing the self-interaction of the polarization tensor. For simplicity, we take the form of $V(M)$ as follows,
\begin{eqnarray}\label{potential}
V(M)&=&(^{*}M_{\mu\nu}M^{\mu\nu})^2=[^{*}(M\wedge M)]^2,
\end{eqnarray}
where $^{*}$ is the Hodge-star operator. As shown in Ref.~\cite{Cai:2015bsa}, this potential shows a global minimum at some nonzero value of $\rho$. Meanwhile, $L(F)$ is the Lagrangian of three classes of Born-Infeld-like nonlinear electrodynamics
\begin{numcases}{L(F)=}
\frac{1}{b} \ln(1+b F),     & LNE \nonumber\\
\frac{1}{b} (1-\sqrt{1+2 b F}),     & BINE \label{action2}\\
-\frac{1}{4 b} (e^{-4 b F}-1).     & ENE \nonumber
\end{numcases}
Here $F\equiv F_{\mu\nu}F^{\mu\nu}$ and $F_{\mu\nu}$ is the nonlinear electromagnetic tensor. As the nonlinear parameter $b$ tends to zero, the Lagrangian $L(F)$ approaches to $F_{\mu\nu}F^{\mu\nu}$. Note that the higher order terms in the parameter $b$ essentially correspond to the higher derivative corrections of the gauge fields. With the same value of $b$, we can discuss the difference in the three types of the holographic dual models with the nonlinear electrodynamics quantitatively. It should be noted that the horizon geometry of nonlinear charged black holes is closed to the horizon of uncharged (Schwarzschild) black hole solution for very large values of $b$~\cite{Hendi:2012zz}, so in this case $L(F)$ can be neglected.
By varying action ~\eqref{action1}, we can get the equations of motion for 2-form field
\begin{eqnarray}\label{eqa1}
\nabla^{\tau}(d M)_{\tau\mu\nu}-m^2 M_{\mu\nu}-J(^{*}M_{\tau\sigma}M^{\tau\sigma})(^{*}M_{\mu\nu})&=&F_{\mu\nu},
\end{eqnarray}
and gauge field
\begin{align}\label{eqa2}
\nabla^{\mu}(\frac{F_{\mu\nu}}{1+b F}+\frac{\lambda^2}{4}M_{\mu\nu})&=0,  ~~~~~\text{LNE} \nonumber\\
\nabla^{\mu}(\frac{F_{\mu\nu}}{\sqrt{1+2 b F}}+\frac{\lambda^2}{4}M_{\mu\nu})&=0,  ~~~~~\text{BINE} \\
\nabla^{\mu}(\frac{F_{\mu\nu}}{e^{4 b F}}+\frac{\lambda^2}{4}M_{\mu\nu})&=0.  ~~~~~~\text{ENE} \nonumber
\end{align}

In what follows, we will work in the probe limit and the background is a 4-dimensional planner Schwarzschild-AdS black hole
\begin{equation}\label{metric1}
ds^2=L^2(-r^{2}f(r) dt^2+\frac{dr^2}{r^2 f(r)}+r^2(dx^2+dy^2)),
\end{equation}
with
\begin{equation}\label{fr}
f(r)=1-\frac{r_{+}^3}{r^3},
\end{equation}
where the $r_{+}$ is the event horizon of the black hole and the Hawking temperature is
\begin{equation}\label{temperature}
T=\frac{3 r_{+}}{4 \pi}.
\end{equation}
In order to study systematically the effects of the $b$ on the holographic ferromagnetic phase transition, we take the following self~-~consistent ansatz with matter fields,
\begin{align}\label{ansatz}
M_{\mu\nu}&=-p(r)dt\wedge dr+\rho(r)dx\wedge dy,\nonumber\\
A_{\mu}&=\phi(r)dt+Bx dy,
\end{align}
where $B$ is a constant magnetic field viewed as the external magnetic field in the boundary field theory.
Thus nontrivial equations of motion read,
\begin{align}\label{eqrhophip}
\rho''+\frac{f'}{f}\rho'-\frac{1}{r^2 f}[m^2+4 J p^2]\rho+\frac{B}{r^2 f}&=0,\nonumber\\
(m^2-\frac{4 J \rho^2}{r^4})p-\phi'&=0,
\end{align}
which are the same form for the three types of nonlinear electrodynamics ~\eqref{eqa2}. For the gauge field $\phi$, however, we obtain the following equations of motion
\begin{align}\label{eqphi3}
(4+8 b (\frac{B^2}{r^4}+\phi'^2)) \phi''+\frac{8 \phi'}{r} (1+(\frac{6 B^2}{r^4}-2 \phi'^2) b) -\lambda^2 \nonumber\\ (p'+\frac{2 p}{r}) [\frac{4 B^2 b}{r^4} (1-2 \phi'^2 b+\frac{B^2 b}{r^4})+(1-2 \phi'^2 b)^2]&=0,  ~~~~\text{LNE} \\
\phi''+\frac{1}{16 B^2 b r+4r^5}(16 B^2 b+8 r^4-32 b r^4 \phi'^2) \phi'\nonumber\\
-\frac{r^3 \lambda^2 (2 p+p'r)}{16 B^2 b+4r^4}(1+\frac{4 B^2 b}{r^4}-4 b \phi'^2)^{3/2}&=0,  ~~~~\text{BINE} \nonumber\\
(4+64 b \phi'^2) \phi''-(p'+\frac{2 p}{r}) \lambda^2 e^{(\frac{8 b B^2}{r^4}-8 b \phi'^2)}+\frac{8 \phi'}{r} (1+\frac{16 B^2 b}{r^4})&=0,  ~~~~\text{ENE} \nonumber
\end{align}
here a prime denotes the derivative with respect to r. Obviously, Eqs.~\eqref{eqrhophip} and ~\eqref{eqphi3} reduce to the standard holographic paramagnetism-ferromagnetism phase transition models discussed in Ref.~\cite{Cai:2015bsa} when $b\rightarrow 0$. In order to solve the nonlinear Eqs.~\eqref{eqrhophip} and ~\eqref{eqphi3} numerically, we should first solve the equation of $\phi'$ and put it into Eq.~\eqref{eqphi3} and get the equation of $p$. And then we need to seek the boundary condition for $\rho$, $\phi$ and $p$ near the black hole horizon $r\rightarrow r_{+}$ and at the spatial infinite $r\rightarrow \infty$. The regularity condition for $\rho(r_{+})$ at the horizon gives the boundary condition $\phi(r_{+})=0$. Near the boundary $r\rightarrow\infty$, the nonlinear equations give the following asymptotic solution for matter fields
\begin{align}\label{asolution}
\rho&=\rho_{+}r^{\Delta_{+}}+\rho_{-}r^{\Delta_{-}}+\cdots+\frac{B}{m^2},\nonumber\\
\phi&=\mu-\frac{\sigma}{r}+\cdots,~p=\frac{\sigma}{m^2 r^2}+\cdots,
\end{align}
with
\begin{align}\label{delta}
\Delta_{\pm}&=\frac{1}{2}\pm\frac{1}{2} \sqrt{1+4 m^2},
\end{align}
where $\rho_{\pm}$, $\mu$ and $\sigma$ are all constants, and $\mu$ and $\sigma$ are interpreted as the chemical potential and charge density in the dual field theory respectively. The coefficient $\rho_{+}$ and $\rho_{-}$ correspond to the source and vacuum expectation value of dual operator in the boundary field theory when $B=0$. Therefore one should set $\rho_{+}=0$ since one wants the condensation to happen spontaneously below a critical temperature. When $B\neq 0$, the asymptotic behavior is governed by external magnetic field $B$.
\begin{figure}
\centering
\includegraphics[width=0.32\textwidth]{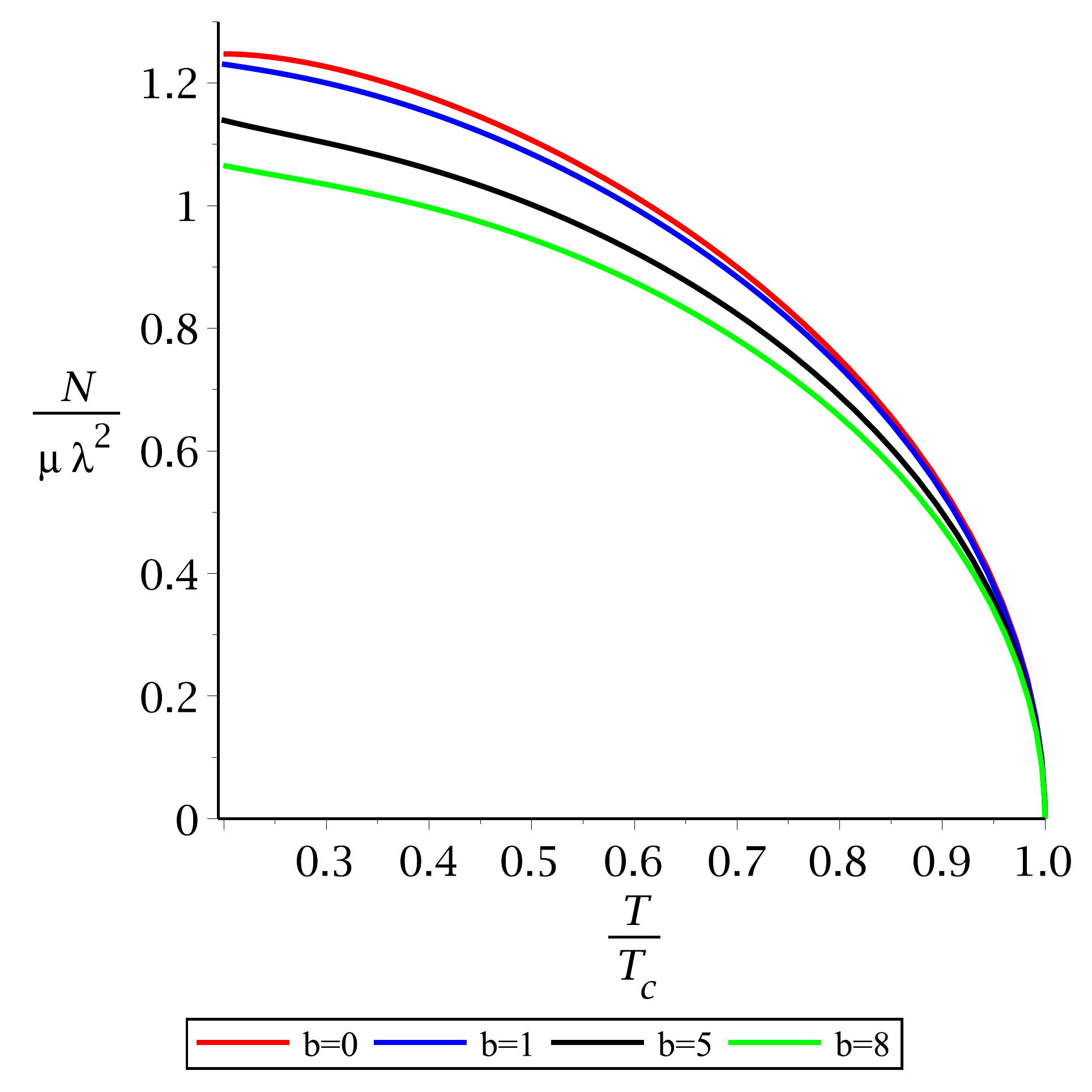}
\includegraphics[width=0.32\textwidth]{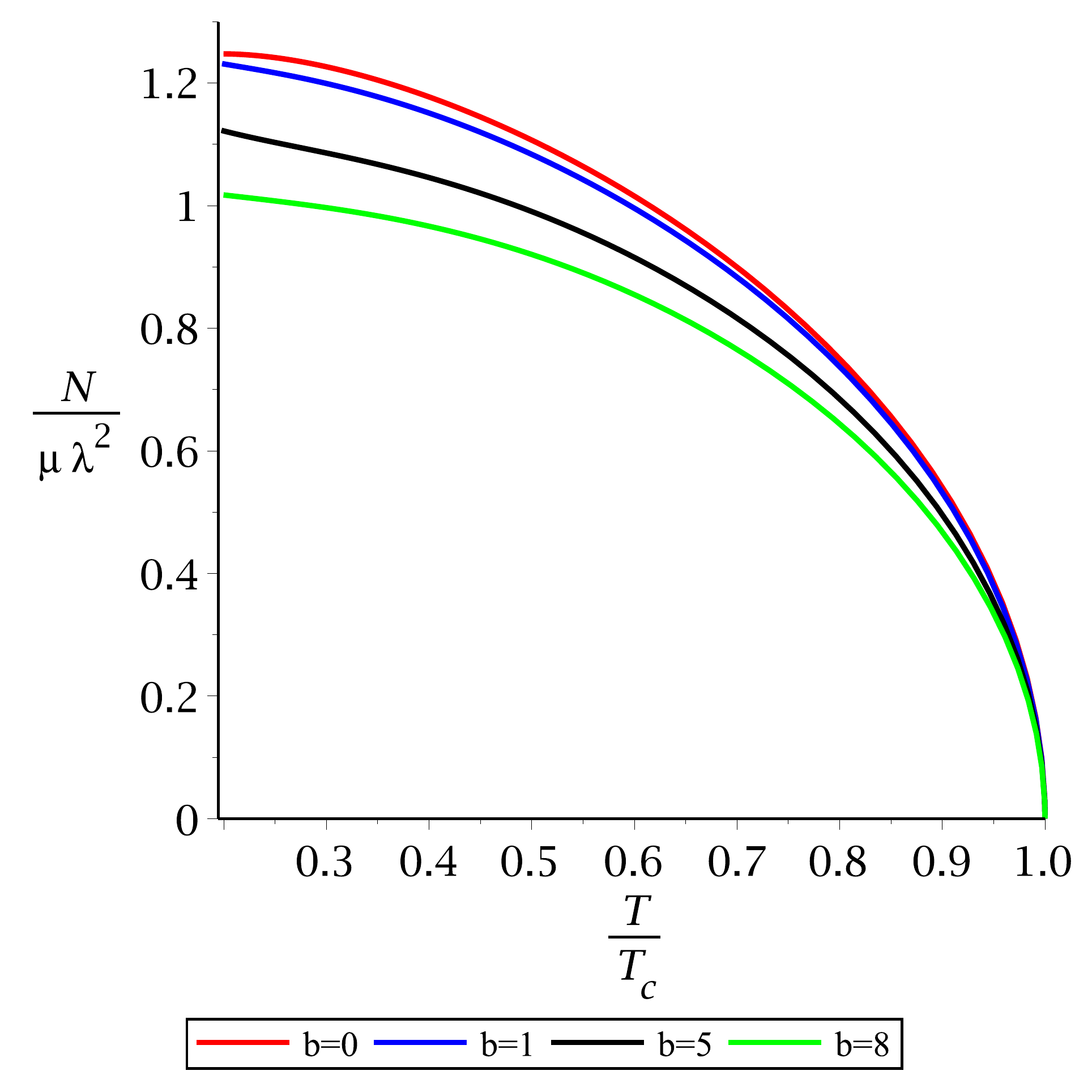}
\includegraphics[width=0.32\textwidth]{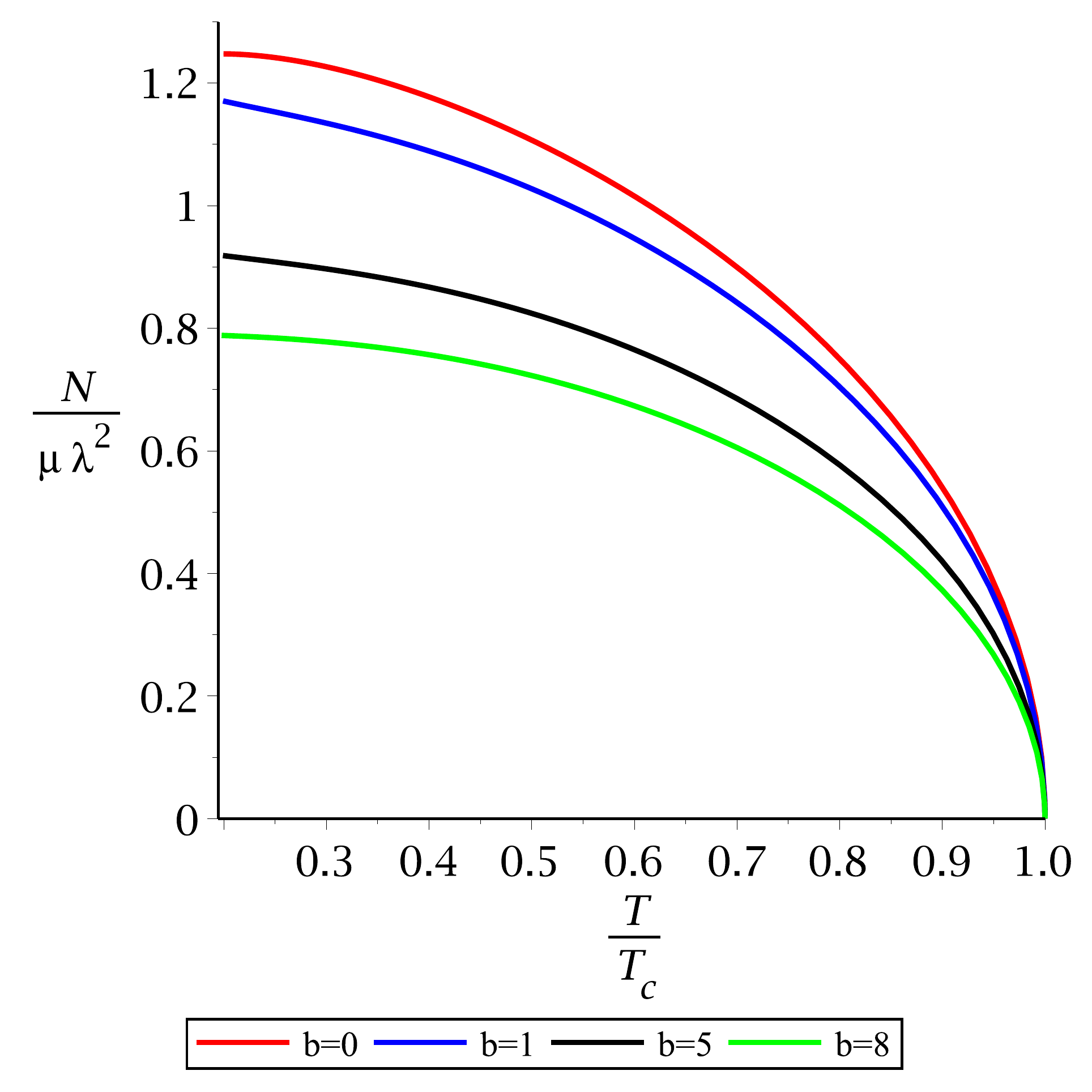}
\includegraphics[width=0.32\textwidth]{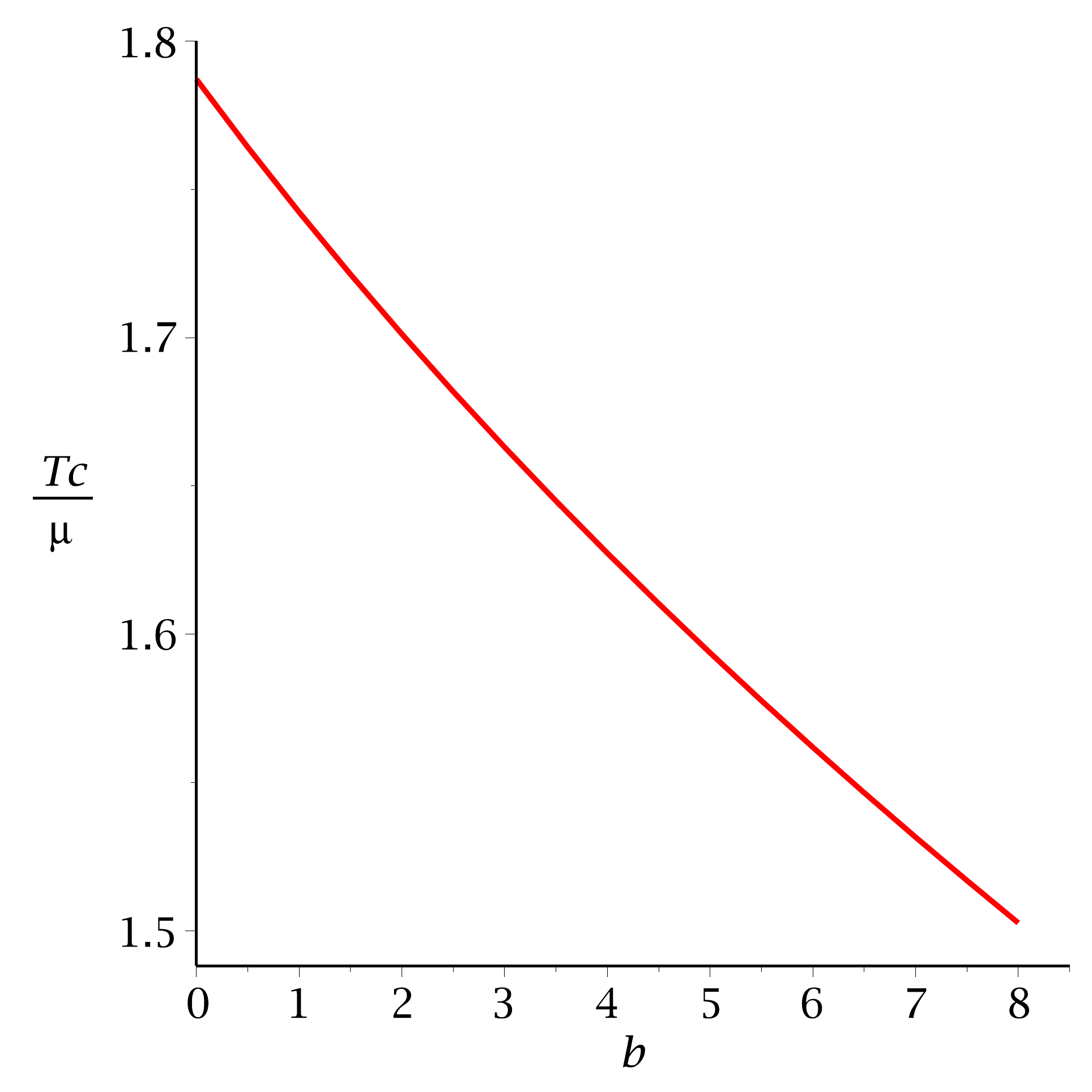}
\includegraphics[width=0.32\textwidth]{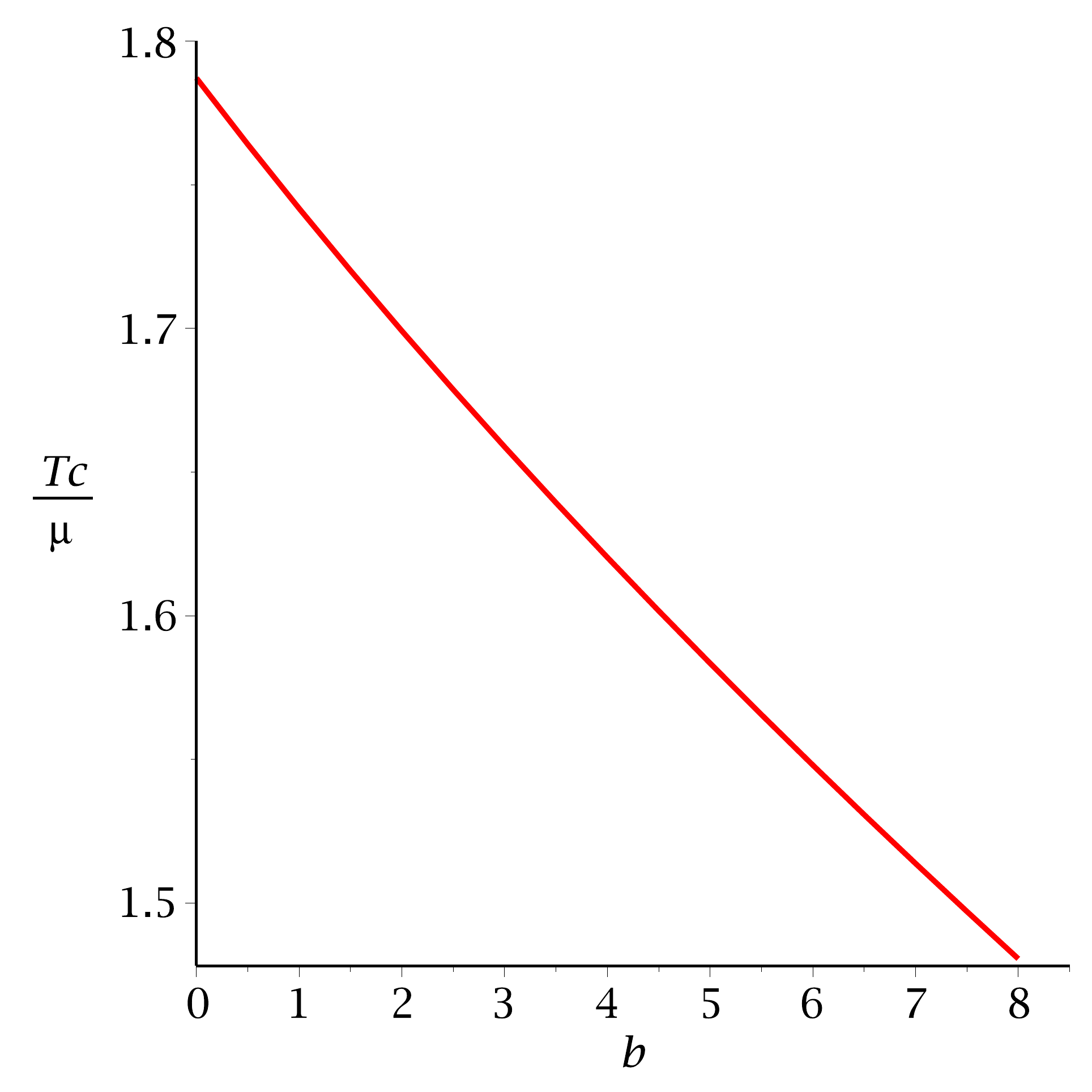}
\includegraphics[width=0.32\textwidth]{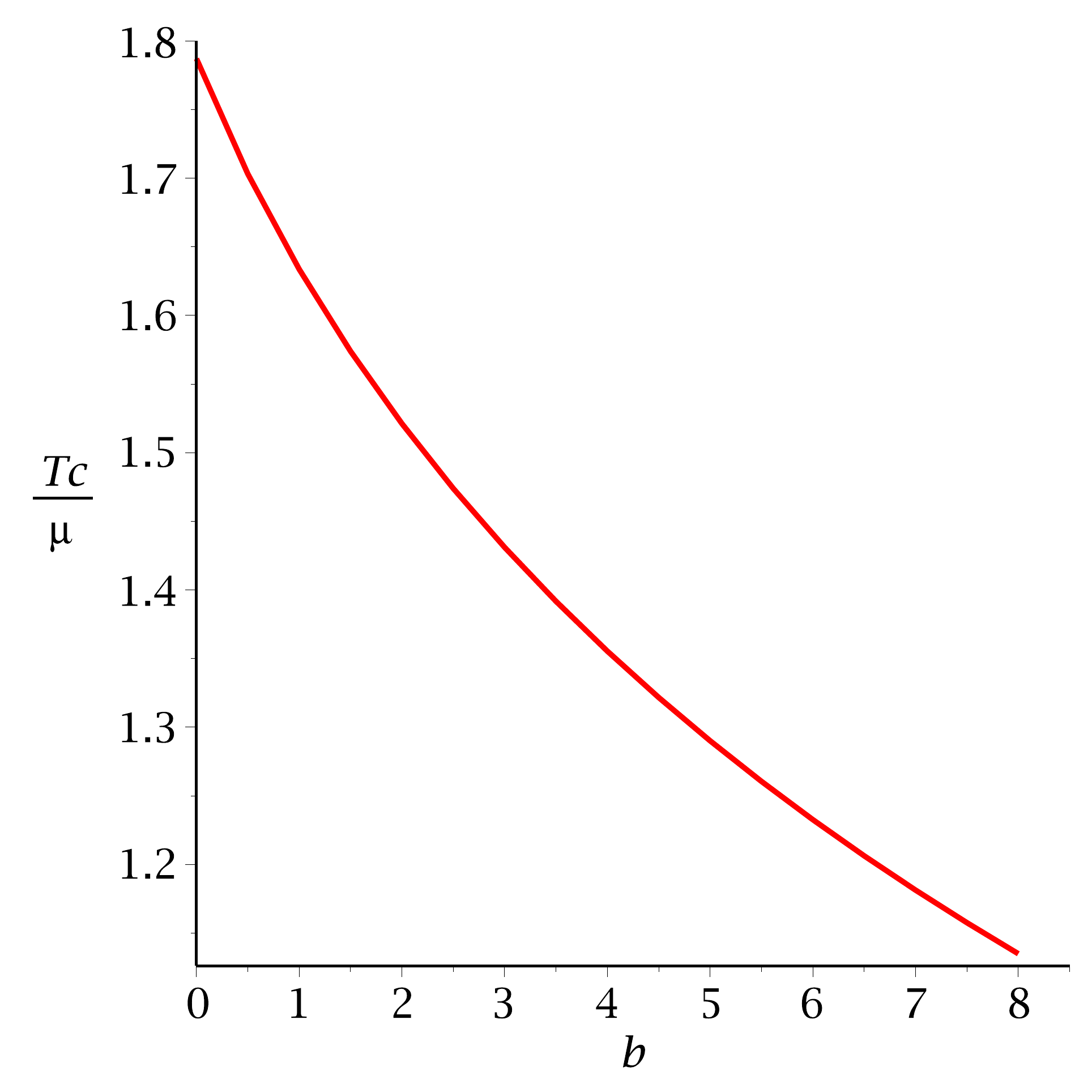}
\caption{The variety of the magnetic moment $N$ and the critical temperature $T_{c}$ with the LNE(left two panels), BINE(middle two panels) and ENE(right two panels) in the presence of nonlinear parameter $b$.} \label{FN}
\end{figure}
\begin{table}
\caption{The critical temperature $T_{c}$ with different values of $b$.}
\label{tab:Tcb}
\centering
\begin{tabular}{p{2.5cm}p{2.5cm}p{2.5cm}p{2.5cm}p{2.5cm}}
\toprule[1pt]
  $b$& $0$ & $1$ & $5$ & $8$ \\ \midrule[0.7pt]
     LNE& $1.7870$ &$1.7422$ &$1.5937$ &$1.5026$ \\
     BINE& $1.7870$ & $1.7417$ & $1.5834$ & $1.4806$ \\
     ENE& $1.7870$ & $1.6337$ & $1.2901$ & $1.1349$\\
\bottomrule[1pt]
\end{tabular}
\end{table}
\begin{table}
\caption{The magnetic moment $N$ with different values of $b$.}
\label{tab:Nb}
\centering
\begin{tabular}{c c c  c c }
\toprule[1pt]
  $b$& $0$ & $1$ & $5$ & $8$ \\ \midrule[0.7pt]
     LNE& $2.9409(1-T/T_{c})^{1/2}$&$2.8529(1-T/T_{c})^{1/2}$ &$2.5177(1-T/T_{c})^{1/2}$ &$2.2908(1-T/T_{c})^{1/2}$ \\
     BINE& $2.9409(1-T/T_{c})^{1/2}$ & $2.8514(1-T/T_{c})^{1/2}$ & $2.4835(1-T/T_{c})^{1/2}$ & $2.2130(1-T/T_{c})^{1/2}$ \\
     ENE& $2.9409(1-T/T_{c})^{1/2}$ & $2.6186(1-T/T_{c})^{1/2}$ & $1.7843(1-T/T_{c})^{1/2}$ & $1.4090(1-T/T_{c})^{1/2}$\\
\bottomrule[1pt]
\end{tabular}
\end{table}

\section{Spontaneous magnetization}
In this paper we work in the grand canonical ensemble where the chemical potential $\mu$ will be fixed. And the expression of magnetic moment as
\begin{equation}\label{eqN}
N=-\lambda^2\int\frac{\rho}{2 r^2} dr.
\end{equation}
Here, we take $J=-1/8$, $m^2=1/8$ and $\lambda=1/2$ as a typical example, which can capture the basic features of the model. In other words, the other choices of the parameters will not qualitatively modify our results. Using the shooting method, we can solve the Eqs.~\eqref{eqrhophip} and~\eqref{eqphi3} numerically and then discuss the effect of the nonlinear electrodynamics on the magnetic moment.

Varying the nonlinearity parameter $b$, we present in the upper half plane of Fig.~\ref{FN} the magnetic moment with the LNE (left two panels), BINE (middle two panels) and ENE (right two panels) as a function of temperature in $d=4$ dimension. It is found that the spontaneous condensate of $\rho$ (corresponding to the magnetic moment) in the bulk in the absence of external magnetic field appears and has similar behavior for different $b$ when the temperature is lower than critical temperature $T_{c}$. Meanwhile, by fitting this curve in the vicinity of critical temperature, we find that the phase transition is a second order one with behavior $N\propto \sqrt{1-T/T_{c}}$ for all cases calculated above. The results are still consistent with one in the mean field theory and have been shown in Tab.~\ref{tab:Nb}. In other words, similar to the case of BINE, the holographic paramagnetism-ferromagnetism transition still exist even we consider the logarithmic and exponential forms of nonlinear electrodynamics.

From the upper right corner of Fig.~\ref{FN}, we observe that the increasing value of the nonlinear parameter $b$ makes the magnetic moment smaller with the ENE, which is similar to the cases of BINE and LNE. It means that the magnetic moment is harder to be formed in the nonlinear electrodynamics, which agrees well with the results given in~\cite{Wu:2016uyj}. In Tab.~\ref{tab:Tcb} and Tab.~\ref{tab:Nb}, we present the critical temperature $T_{c}$ and the behaviors of these condensation curves near $T\sim T_{c}$. It is easy to find that as $b$ increases the critical temperature decreases for each nonlinear electrodynamic, which is exhibited in the lower half plane of Fig.~\ref{FN} and agrees well with the finding in the upper half plane of Fig.~\ref{FN}. This behavior has been seen for the holographic superconductor in the background of a Schwarzschild~-~AdS black hole, where the three types of typical nonlinear electrodynamics make scalar condensation harder to form~\cite{Zhao:2012cn}. At the same time, the dependence of the magnetic moment and the critical temperature on the nonlinear parameter is similar to that on the Gauss-Bonnet term in the holographic superconductor, i.e., the higher curvature corrections make condensation harder to form. Therefore, we conclude that the ENE, BINE and LNE corrections to usual Maxwell field and the curvature corrections share some similar features for the condensation of the massive 2-form field $\rho$.

On the other hand, comparing with the curves for the magnetic moment in the three types of the nonlinear electrodynamics considered here, we find that the value of magnetic moment with ENE is smaller than ones in the BINE and LNE cases for the fixed value of nonlinear parameter $b$ (except the case of $b=0$, i.e., the usual Maxwell electrodynamics), which means that the magnetic moment is more difficult to be developed in the exponential form of nonlinear electrodynamics. This is also in good agreement with the results shown in Tab.~\ref{tab:Tcb} and in lower half plane of Fig.~\ref{FN}, where the critical temperature $T_c$ for the condensate of $\rho$ with the ENE is smaller than ones in the BINE and LNE cases for the fixed value of $b$.

\section{The response to the external magnetic field}
Let us turn on the external field to examine the response to magnetic moment $N$. This can be described by magnetic susceptibility density $\chi$, defined as
\begin{equation}\label{definechi}
\chi=\lim_{B\rightarrow0}\frac{\partial N}{\partial B}.
\end{equation}
In the high temperature region $T>T_{c}$, the ferromagnetic material is in a paramagnetic phase whose magnetic moments are randomly distributed. So the susceptibility obeys the Curie-Weiss law
\begin{align}\label{chi}
\chi&=\frac{C}{T+\theta}, T>T_{c}, \theta<0,
\end{align}
where C and $\theta$ are two constants. Note that a significant difference between the antiferromagnetism and paramagnetism can be seen from the magnetic susceptibility. In the paramagnetic phase of antiferromagnetic material and paramagnetic material, the magnetic susceptibility also obeys the Curie-Weiss law, but the constant $\theta$ in Eq.~\eqref{chi} is positive and zero, respectively.
\begin{figure}
\centering
\includegraphics[width=0.32\textwidth]{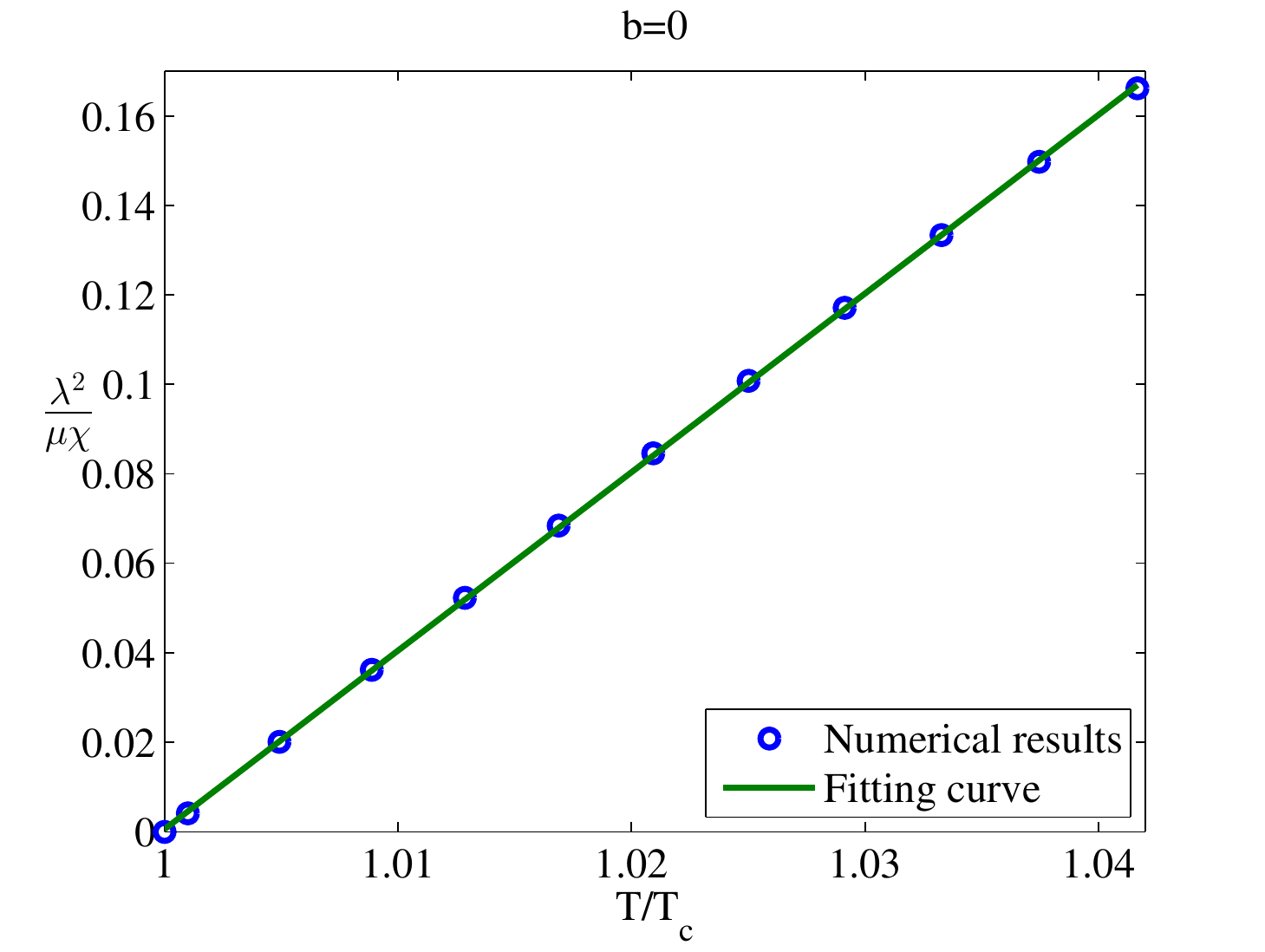}
\includegraphics[width=0.32\textwidth]{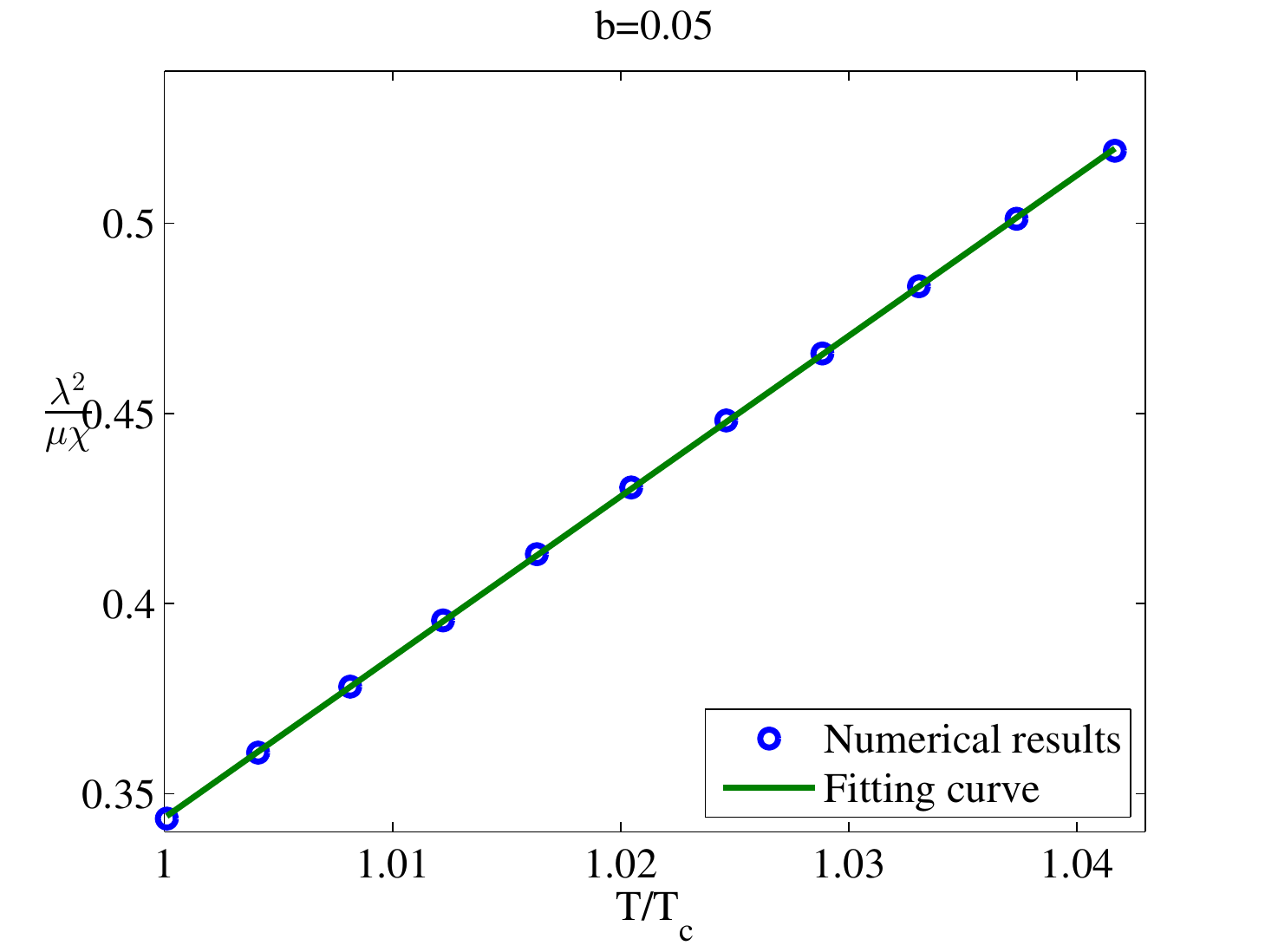}
\includegraphics[width=0.32\textwidth]{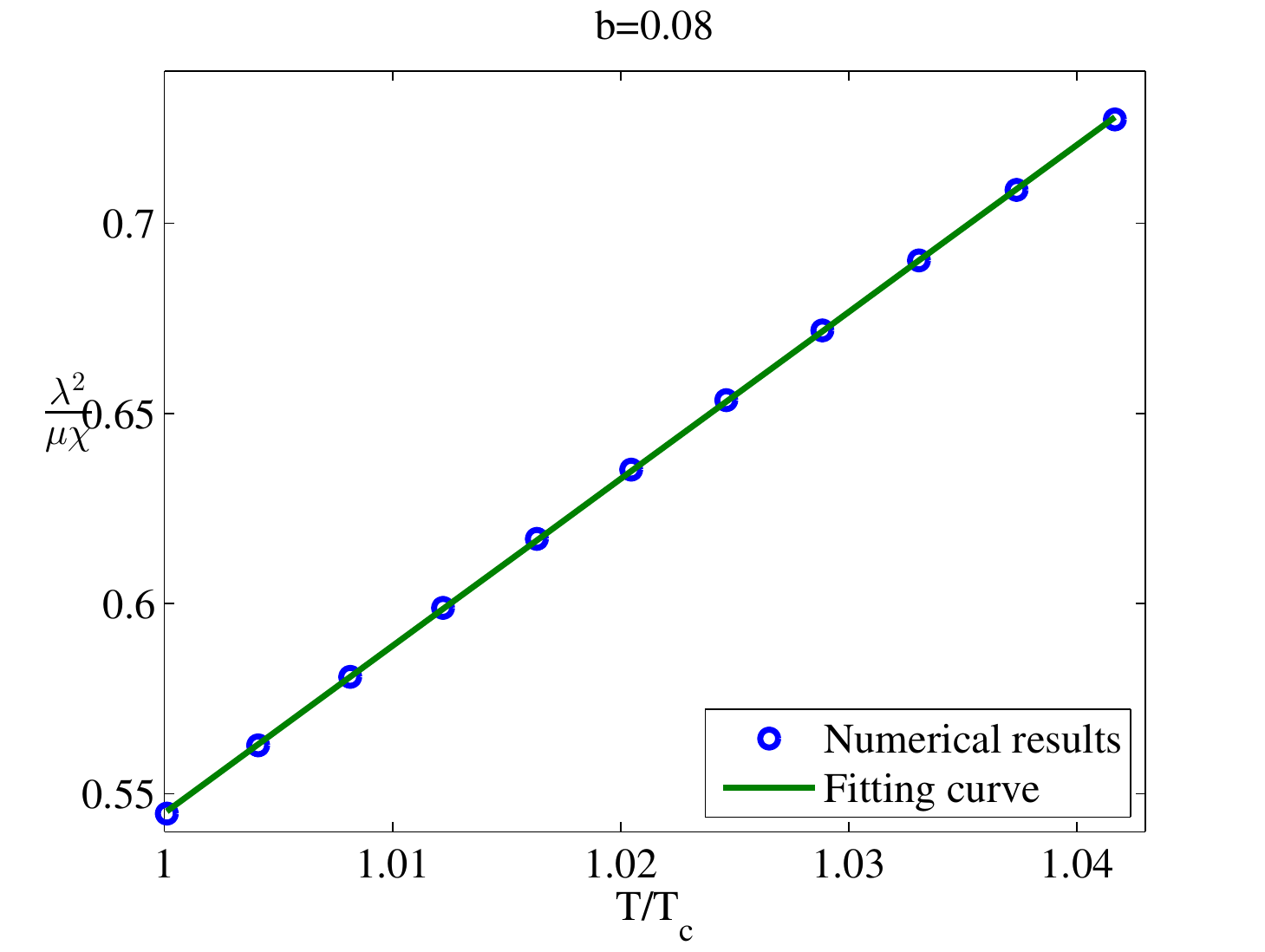}
\includegraphics[width=0.32\textwidth]{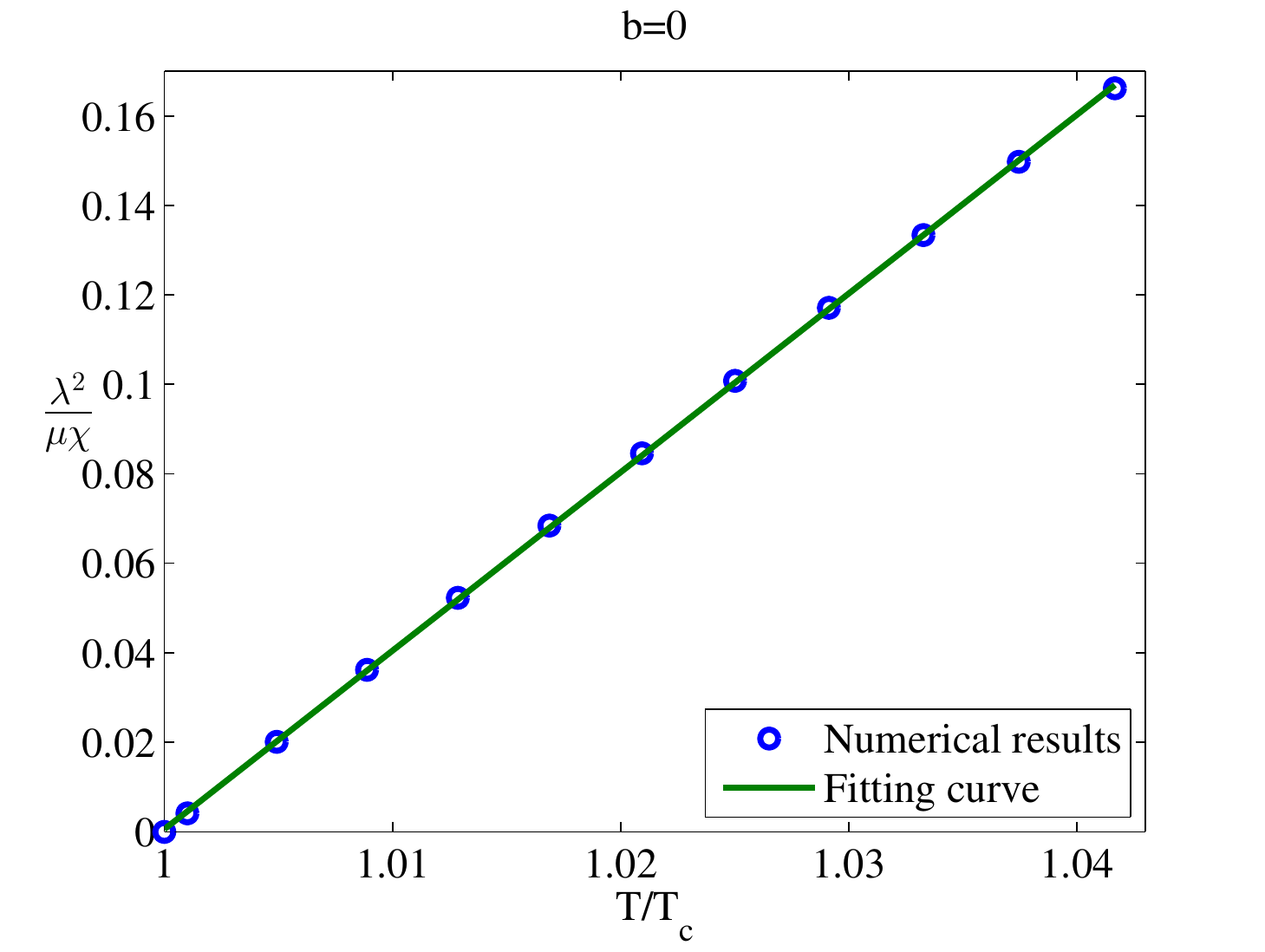}
\includegraphics[width=0.32\textwidth]{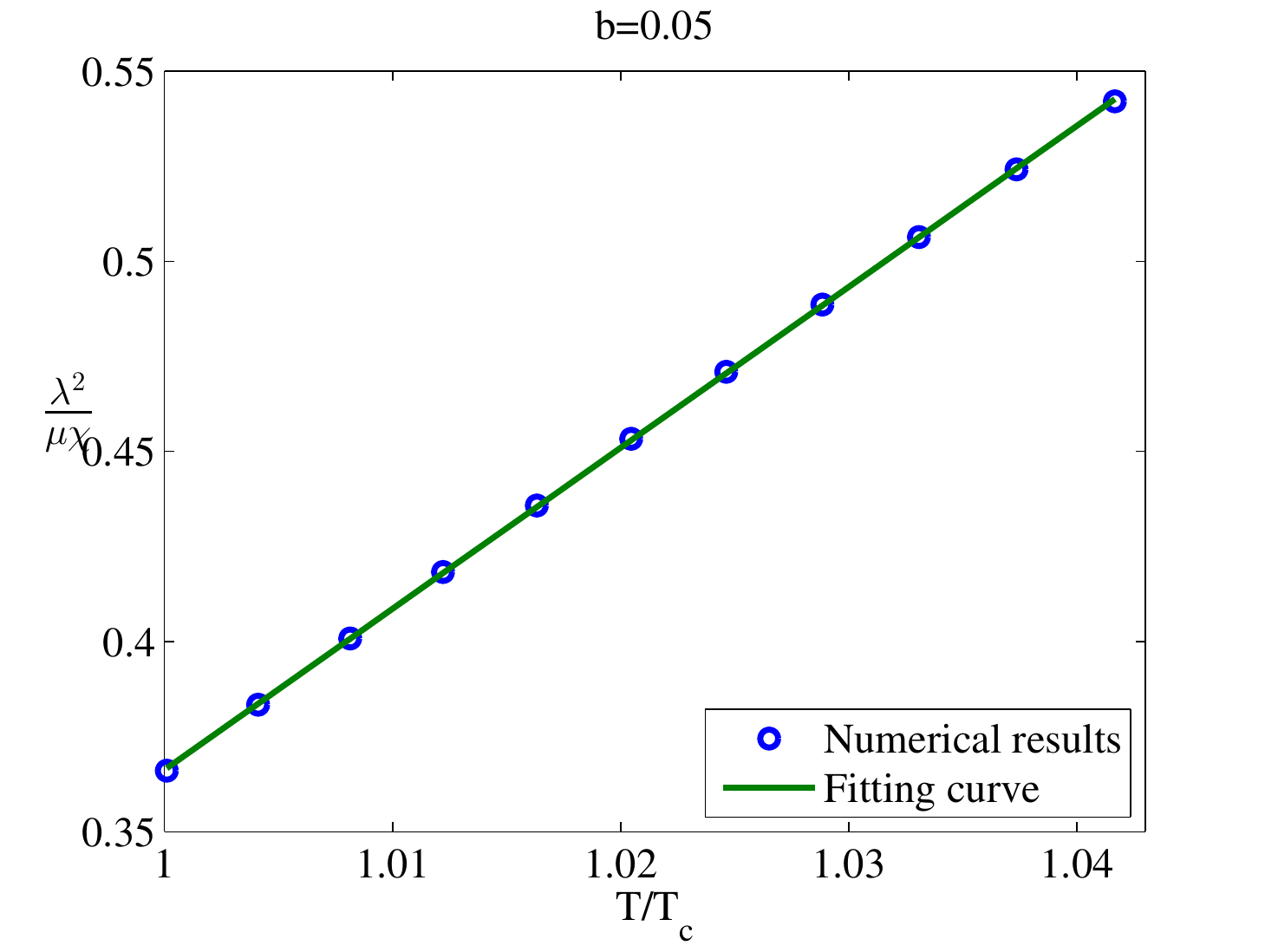}
\includegraphics[width=0.32\textwidth]{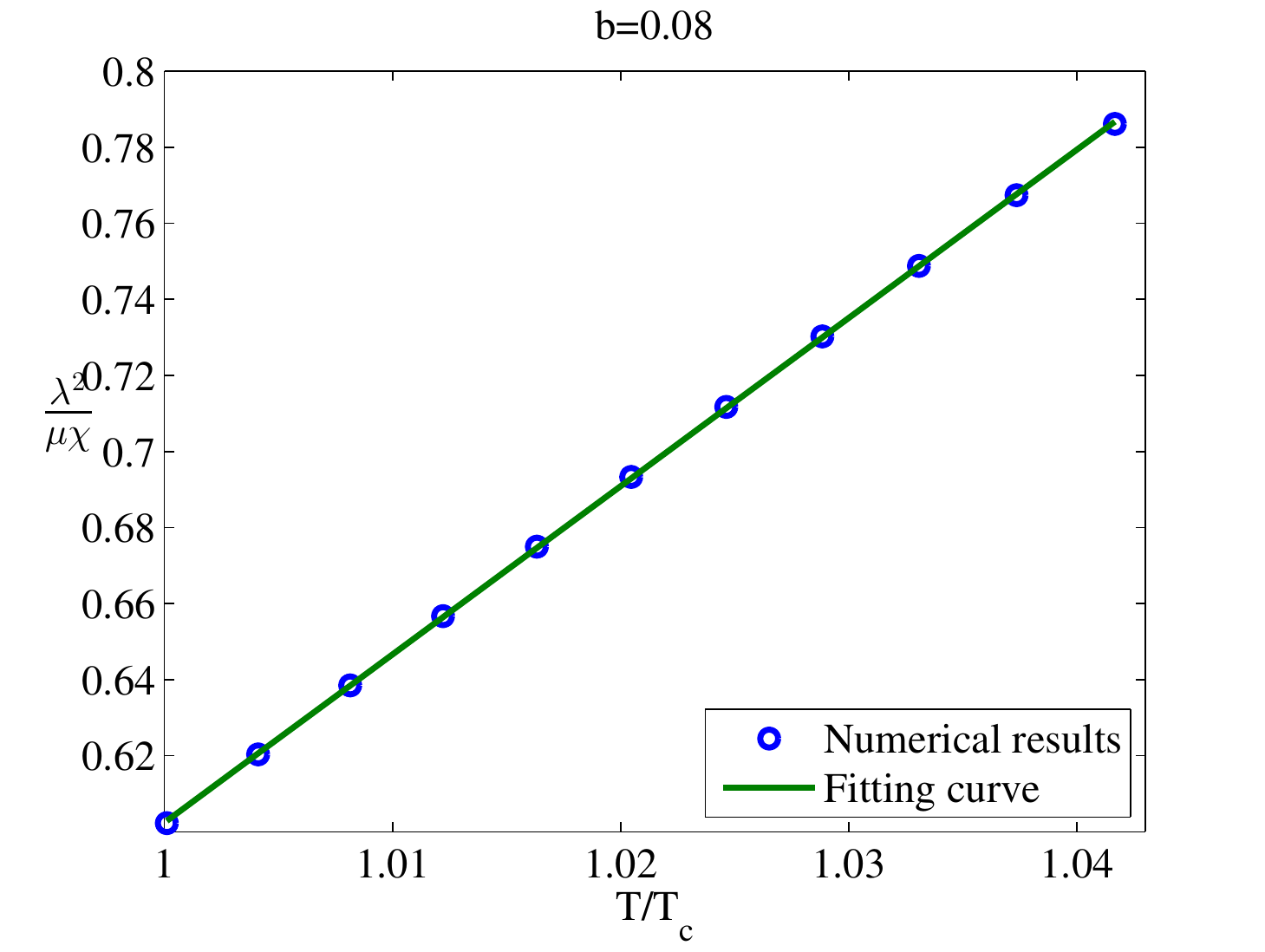}
\includegraphics[width=0.32\textwidth]{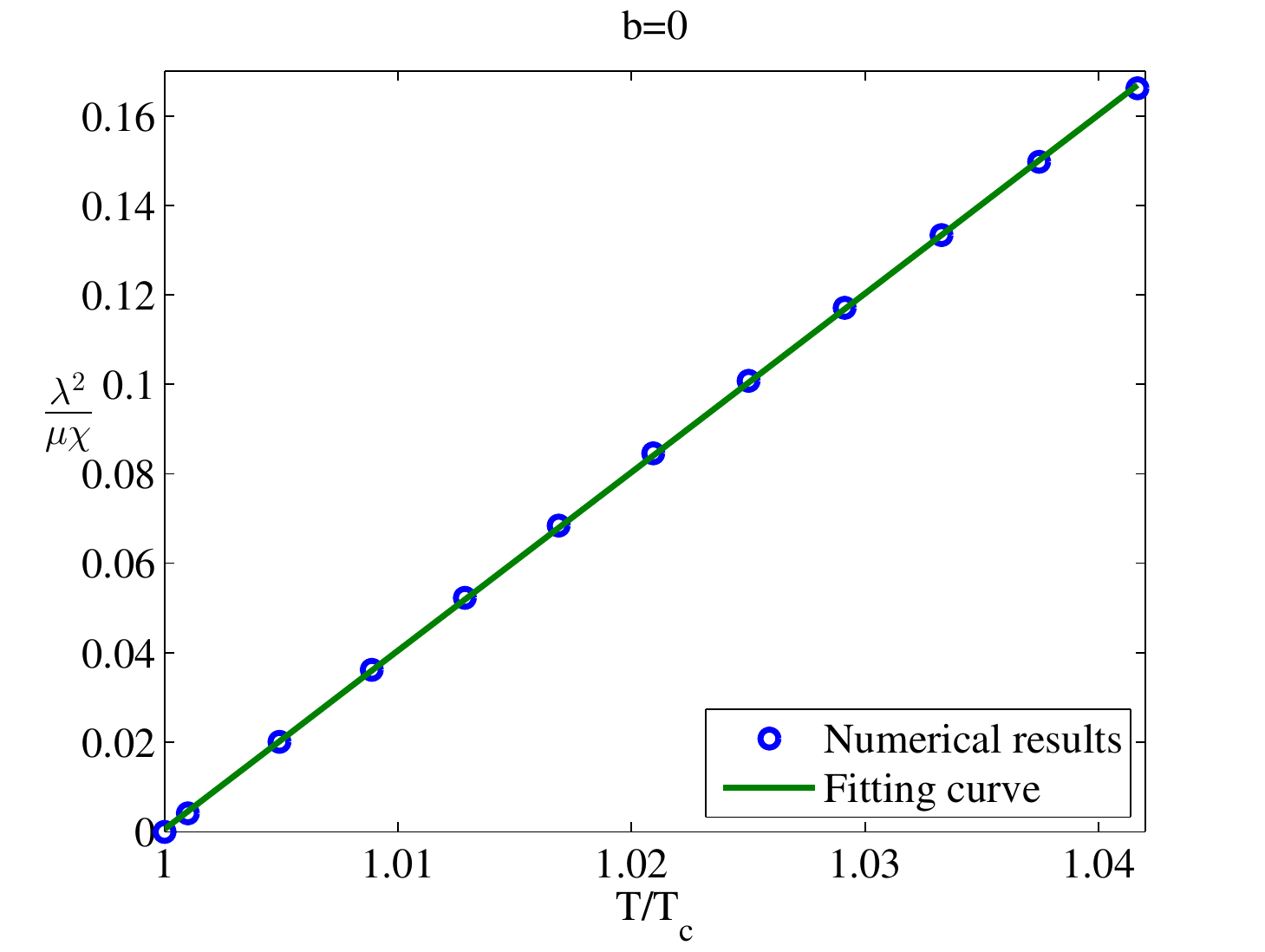}
\includegraphics[width=0.32\textwidth]{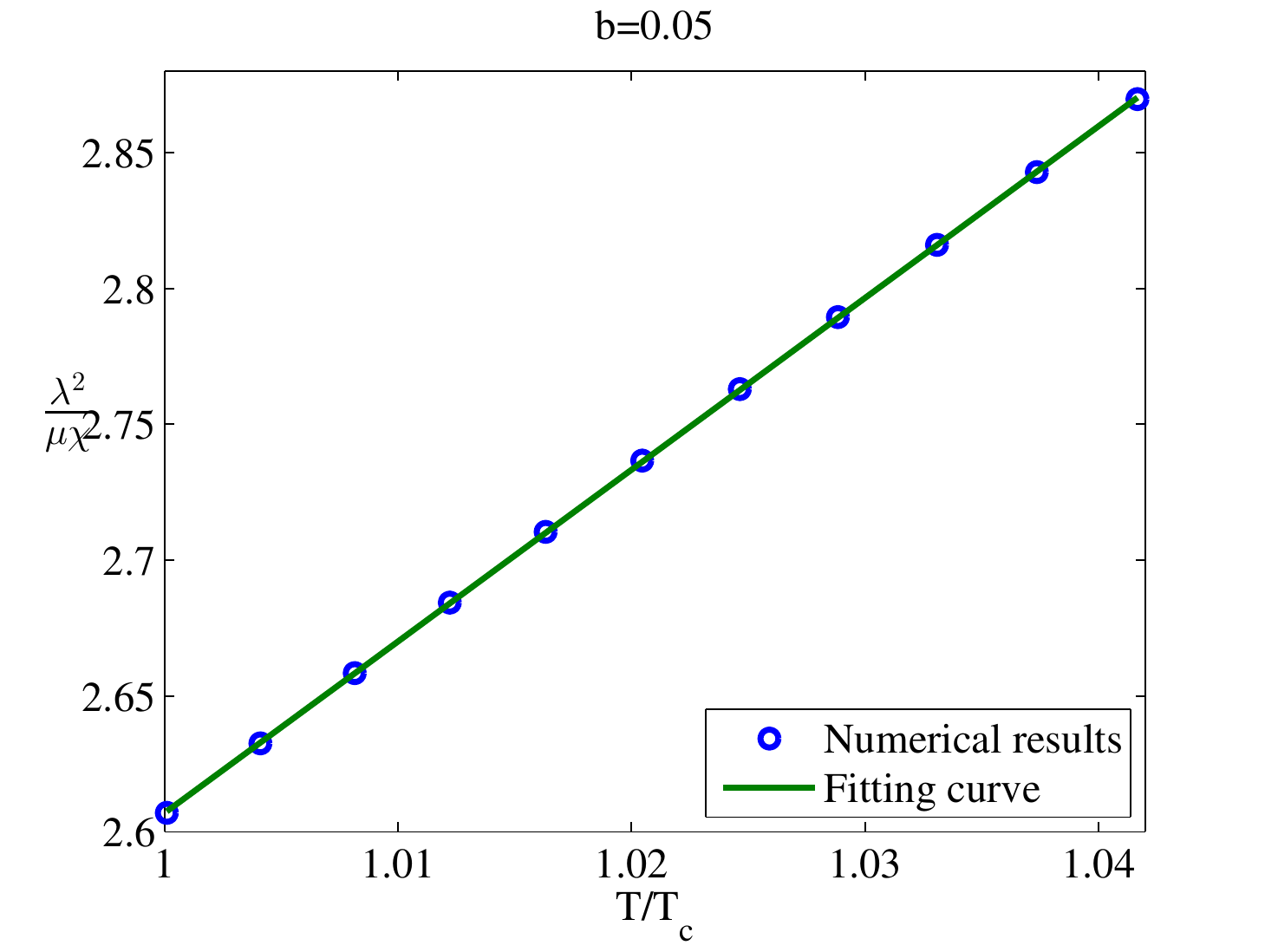}
\includegraphics[width=0.32\textwidth]{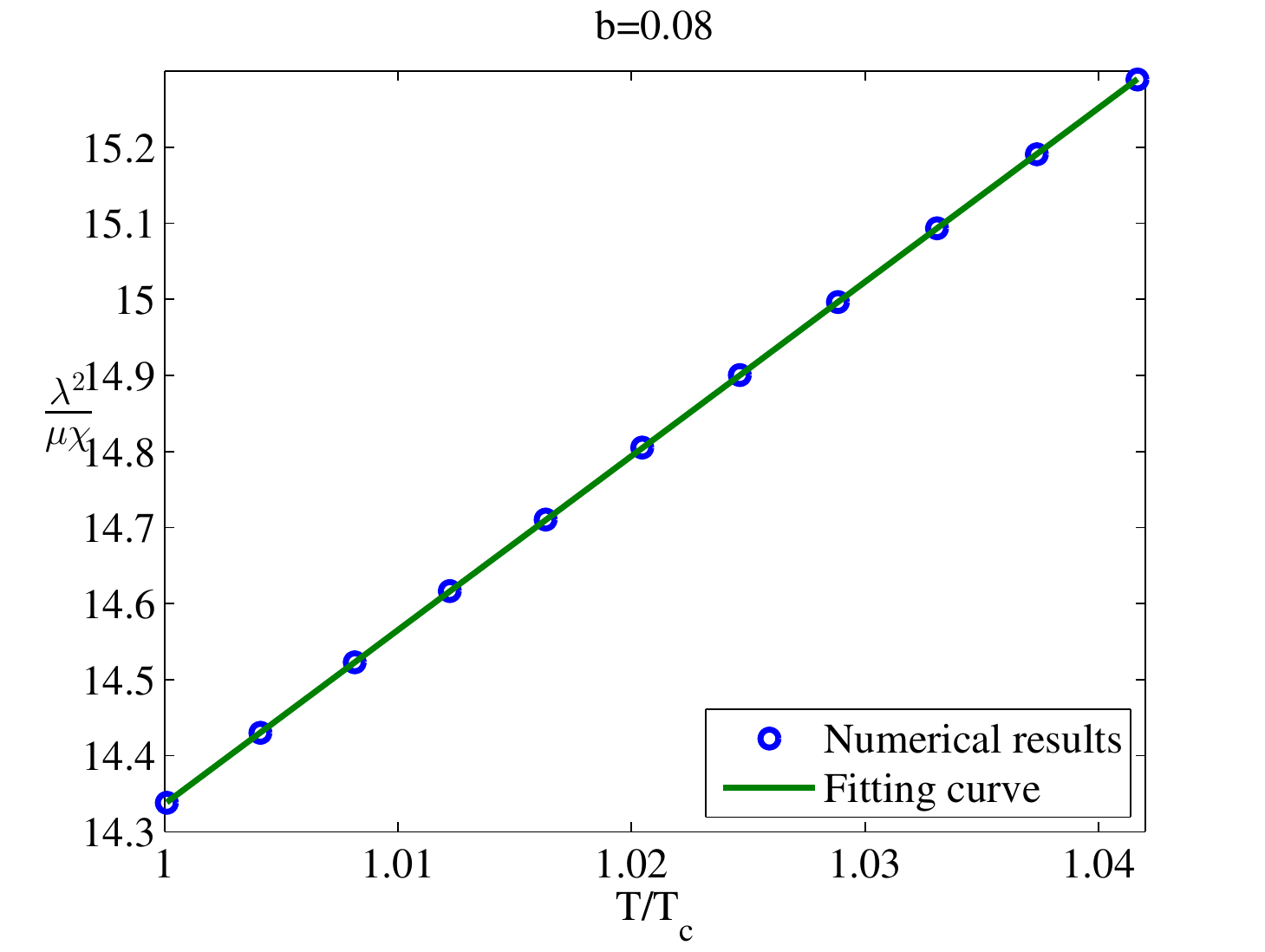}
\caption{The magnetic susceptibility as a function of temperature with the BINE (top three panels), LNE (middle three panels), and ENE (bottom three panels) in the presence of nonlinear parameter $b$} \label{chib}
\end{figure}
For the three types of nonlinear electrodynamics, Fig.~\ref{chib} shows the magnetic susceptibility as a function of temperature by solving Eq.~\eqref{definechi} with $b=0$, $0.05$, $0.08$. In the paramagnetic phase for all cases considered here, we observe that the magnetic susceptibility increases when the temperature is lowered for the fixed nonlinearly parameter $b$. Moreover, the magnetic susceptibility satisfies the Curie-Weiss law of the ferromagnetism near the critical temperature whether $b=0$ or not. Concretely, the results have been presented in Tab.~\ref{tab:chib} for the chosen model parameters. It is easy to see that coefficient in front of $\frac{T}{T_{c}}$ for $\frac{1}{\chi}$ increases with the increasing $b$, which meets well with the discovery in Fig.~\ref{chib}. However, the absolute value of $\frac{\theta}{\mu}$ will decrease when the Born-Infeld scale parameter $b$ increases.
\begin{table}
\caption{The magnetic susceptibility $\chi$ with different values of $b$.}
\label{tab:chib}
\centering
\begin{tabular}{c c c  c c }
\toprule[1pt]
  & $b$& $0$ & $0.05$ & $0.08$ \\ \midrule[0.7pt]
     \multirow{2}*{LNE}  & $\lambda^2/\chi \mu$ & $3.9906(T/T_{c}-1)$ &$4.2328(T/T_{c}-0.91)$ &$4.4227(T/T_{c}-0.86)$ \\
                         & $\theta/\mu$ & $-1.7870$ &$-1.6303$ &$-1.5404$ \\
     \multirow{2}*{BINE}  & $\lambda^2/\chi \mu$ & $3.9906(T/T_{c}-1)$ &$4.2227(T/T_{c}-0.92)$ &$4.3911(T/T_{c}-0.88)$ \\
                          & $\theta/\mu$ & $-1.7870$ &$-1.6395$ &$-1.5620$ \\
     \multirow{2}*{ENE}  & $\lambda^2/\chi \mu$ & $3.9906(T/T_{c}-1)$ & $6.3220(T/T_{c}-0.5877)$ & $22.8769(T/T_{c}-0.3733)$ \\
                         & $\theta/\mu$ & $-1.7870$ & $-1.0448$ & $-0.6616$ \\
\bottomrule[1pt]
\end{tabular}
\end{table}
On the other hand, from Fig.~\ref{chib} and Tab.~\ref{tab:chib} we can see the value of coefficient in front of $\frac{T}{T_{c}}$ for $\frac{1}{\chi}$ of the ENE is larger than that of BINE and LNE for the fixed value of $b$ (except the case of $b=0$, i.e., the usual Maxwell electrodynamics). Comparing the cases of BINE and LNE, however, the absolute value of $\frac{\theta}{\mu}$ for ENE is smaller.
\begin{figure}
\centering
\includegraphics[width=0.32\textwidth]{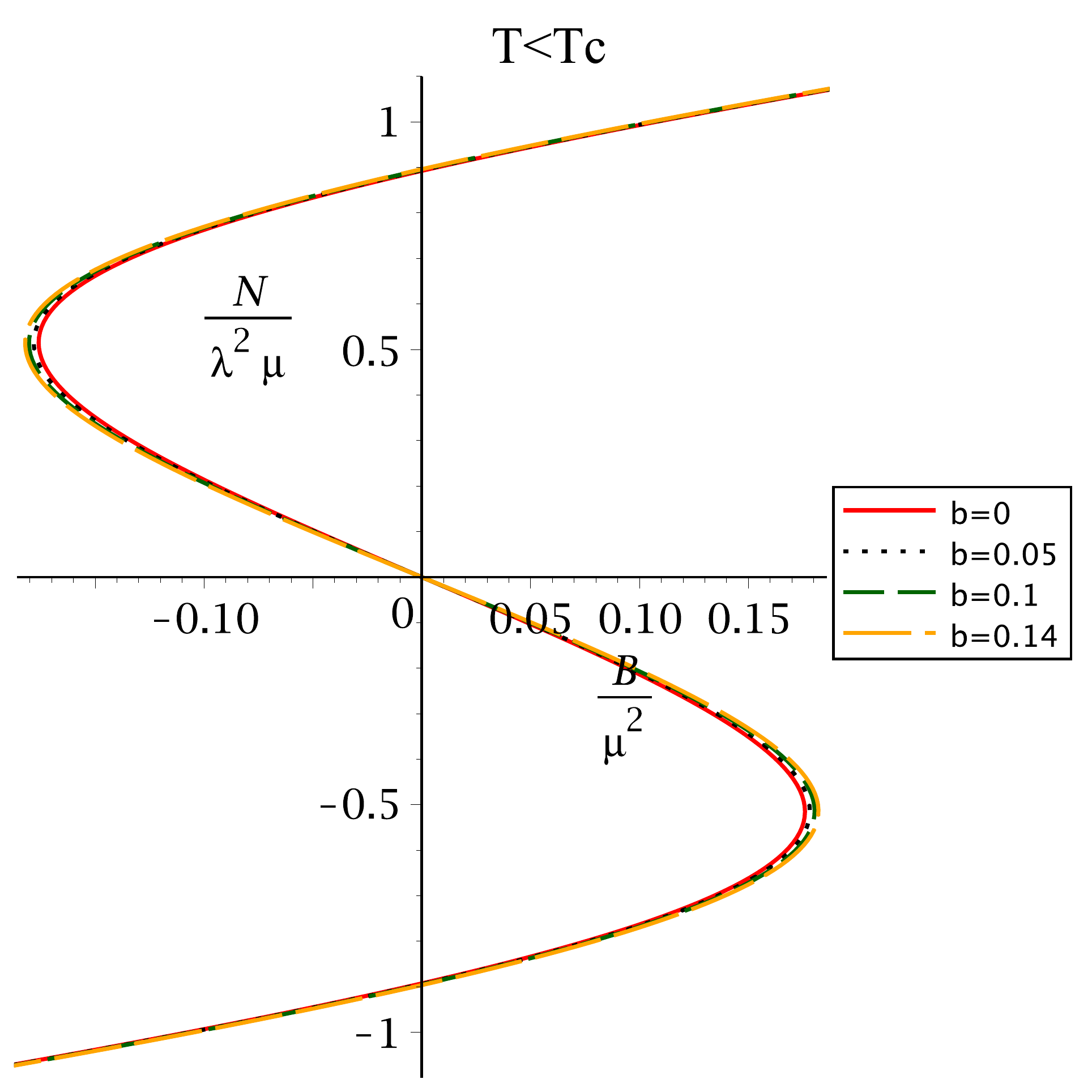}
\includegraphics[width=0.32\textwidth]{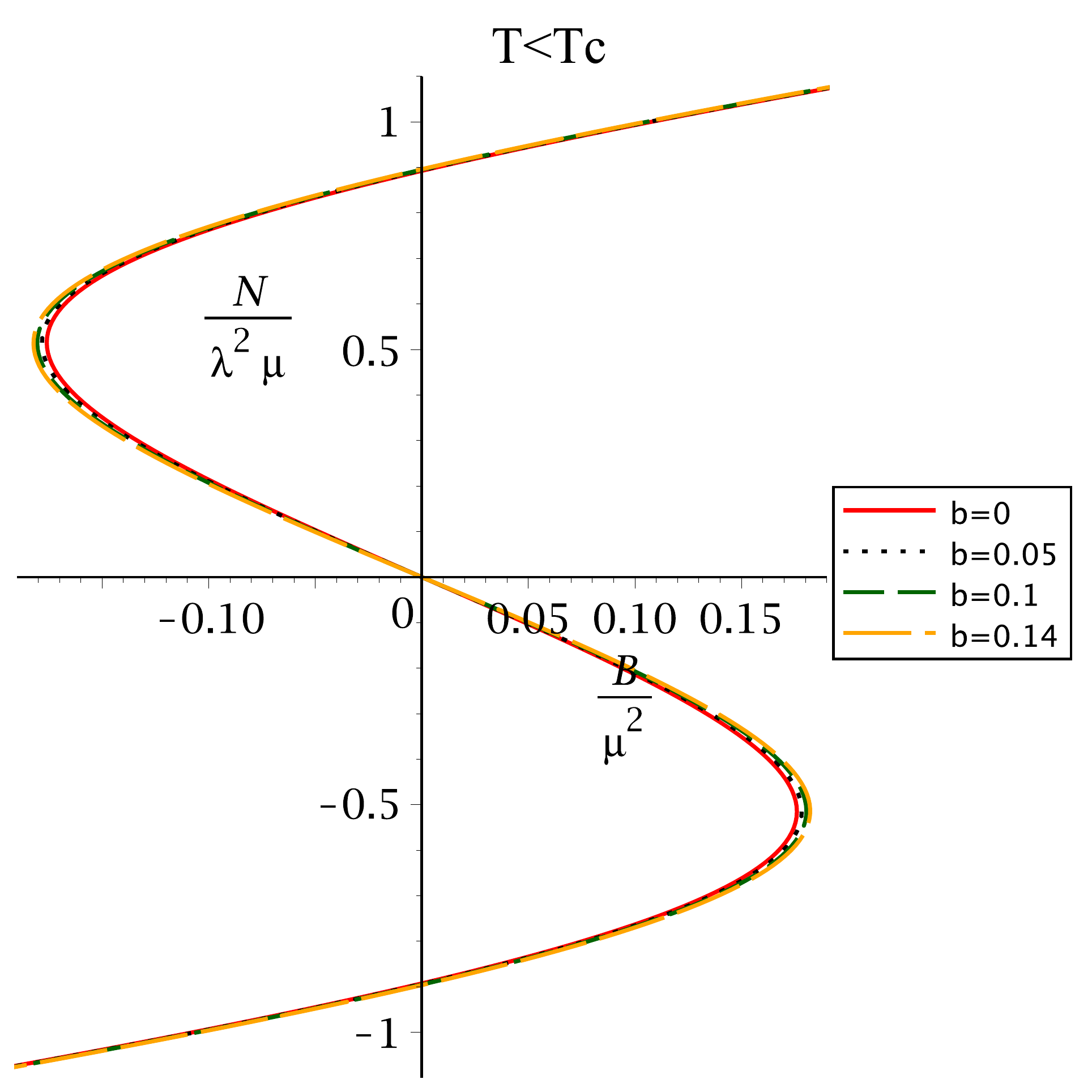}
\includegraphics[width=0.32\textwidth]{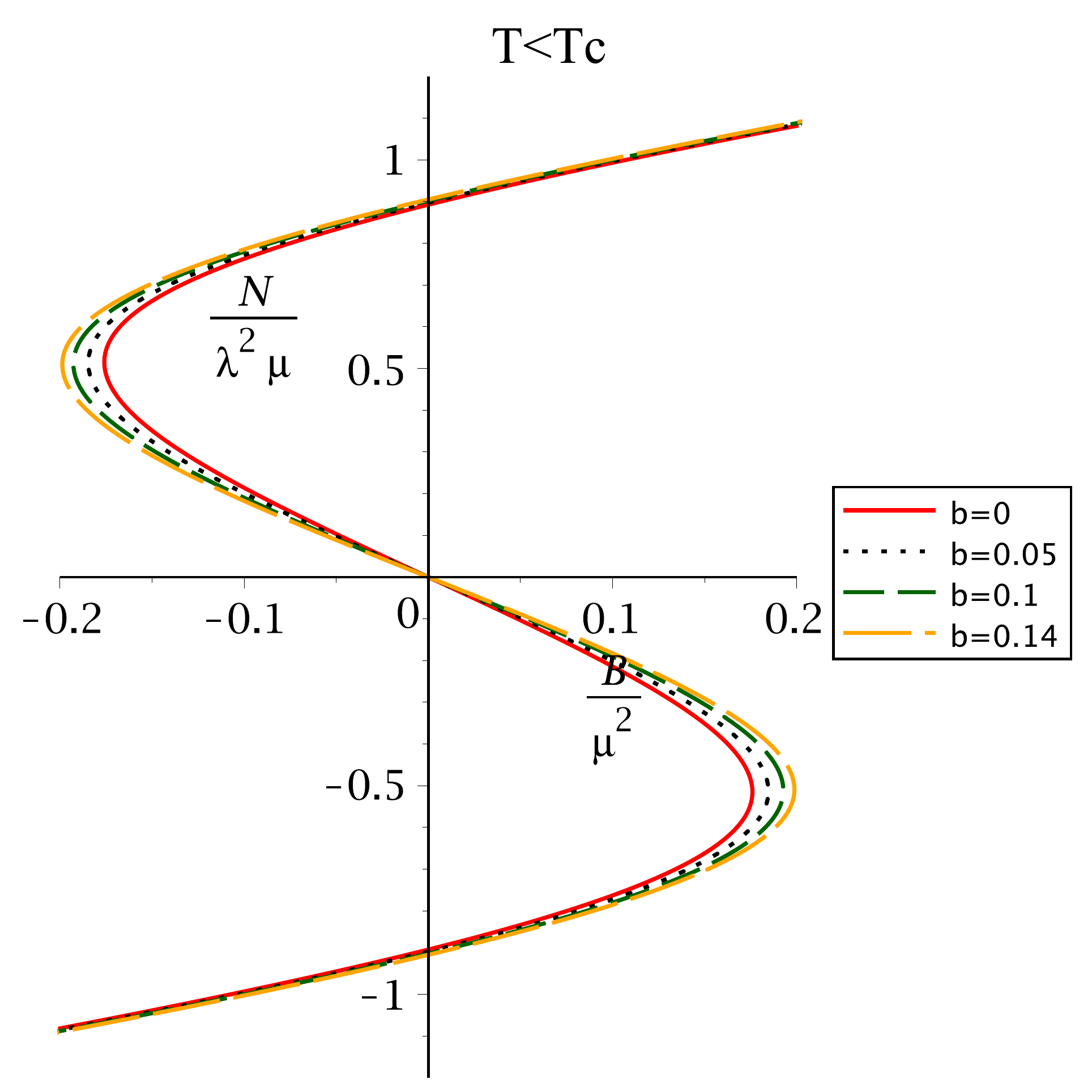}
\caption{The magnetic moment with respect to external magnetic field B in lower temperature with the LNE(left panel), BINE(middle panel) and ENE(right panel) in the presence of nonlinear parameter $b$.} \label{bBN}
\end{figure}
In the plot of Fig.~\ref{bBN}, we show that the magnetic moment with respect to external field $B$ in region of $T<T_{c}$( i.e., $T=0.89 T_{c}$) with different parameter $b$. And from the each line in Fig.~\ref{bBN}, we see that the magnetic moment is not single valued when the external magnetic field continuously changes between $-B_{max}$ and $B_{max}$ periodically. Thus a hysteresis loop in the single magnetic domain will be obtained and the nonlinear parameter $b$ has an effect on it quantitatively. Along the horizontal direction(the magnetic moment has been taken a same value), one need a larger external field as the nonlinear parameter $b$ increases. In other words, the nonlinear electrodynamics makes the periodicity of hysteresis loop bigger which is different from the effect of Lifshitz dynamical exponent $z$ on it. Particularly, for the case of ENE, whose effect on the periodicity of hysteresis loop is more noticeable. However, all the curves will overlap once the value of the magnetic field exceeds the maximum that corresponding to the case of $b=0.14$, which can be seen from Fig.~\ref{bBN}.

\section{Summary and discussion}
Sum up, we have investigated systematically holographic paramagnetism-ferromagnetism phase transitions in the presence of three kinds of typical Born~-~Infeld~-~like nonlinear electrodynamics correction to the Maxwell electrodynamics in 4-dimension Schwarzschild-AdS black hole spacetime, and obtained the effect of the nonlinear parameter $b$ on the holographic paramagnetism-ferromagnetism phase transition. Considering that these nonlinear generalizations essentially imply the higher derivative corrections of the gauge fields, this study may help to understand the influences of the $1/N$ or $1/\lambda$ corrections on the holographic dual model. In the probe limit, comparing the exponential form of nonlinear electrodynamics(ENE) with the Born-Infeld nonlinear electrodynamics(BINE) and logarithmic form of nonlinear electrodynamics(LNE), in the black hole background, we found it has stronger effects on critical temperature and the magnetic moment. Furthermore, we observed that for all three types of the nonlinear electrodynamics considered here the higher nonlinear electrodynamics correction term can make the condensation harder form and result in the decreases of critical temperature and magnetic moment in the absence of magnetic field. This behavior is similar to that of the holographic superconductor in the background of a Schwarzschild-AdS black hole, where the three kinds of nonlinear electrodynamics make scalar condensation harder form and result in the larger deviations from the universal value $\omega_{g}/T_{c}\approx8$ for the gap frequency. In the vicinity of the critical point, however, the behavior of the magnetic moment is always as $(1-T/T_{c})^{1/2}$, which is independent of the explicit form of the nonlinear electrodynamics, i.e., the ENE, BINE, LNE correction terms do not have any effect on the relationship. Meanwhile, it is in agreement with the result from mean field theory. Moreover, in the presence of the external magnetic field, the inverse magnetic susceptibility as $T\sim T_{c}$ behaves as $C/(T+ \theta)$, $(\theta<0)$ in all cases, which satisfies the Cure-Weiss law. Yet both the constant $C$ and the absolute value of $\theta$ decrease with the increasing nonlinear parameter $b$. Furthermore, we have observed the hysteresis loop in the single magnetic domain when the external field continuously changes between the maximum and minimum values periodically with $b$. The increase of the nonlinear parameter $b$ could result in extending the period of the external magnetic field. Especially, the effect of the exponential form of nonlinear electrodynamics on the periodicity of hysteresis loop is more noticeable.

Note that in this paper we just investigate the influence of the three kinds of nonlinear electrodynamics on paramagnetism-ferromagnetism phase transition. It would be of interest to generalize our study to holographic paramagnetism-antiferromagnetism model. Work in this direction will be reported in the future.

\section*{Acknowledgments}
We would like to thank Prof. R. G. Cai and Dr. R. Q. Yang for their helpful discussions and comments. This work is supported by the National Natural Science Foundation of China (Grant Nos. 11175077, 11575075).

\end{document}